\documentclass[conference]{IEEEtran}
\IEEEoverridecommandlockouts
\usepackage[switch]{lineno}
\usepackage{lipsum}
\usepackage{amsmath,amsfonts}
\usepackage{algorithmic}
\usepackage{textcomp}
\usepackage{xcolor}
\usepackage{subfiles}
\usepackage{caption}
\usepackage{subcaption}
\usepackage[symbol]{footmisc}

\usepackage[ruled, lined, linesnumbered, commentsnumbered, longend]{algorithm2e}
\usepackage{array}
\usepackage{multirow}
\usepackage{balance}
\usepackage{flushend}
\usepackage{enumitem} 
\usepackage{tikz}
\usepackage{pifont}% http://ctan.org/pkg/pifont
\usepackage{array}
\usepackage[font=small,labelfont=bf]{caption}
\usepackage[mathscr]{euscript}
 
\newcolumntype{P}[1]{>{\centering\arraybackslash}p{#1}}

\def\BibTeX{{\rm B\kern-.05em{\sc i\kern-.025em b}\kern-.08em
    T\kern-.1667em\lower.7ex\hbox{E}\kern-.125emX}}

\usepackage{amsmath}
\usepackage{float}
\usepackage{booktabs}
\usepackage{natbib}
\begin{document}
%\onecolumn
%\firstpage{1}

\title
%[Deep-Learning Framework for COVID-19 Testing]
{A Deep-Learning Framework for Improving COVID-19 CT Image Quality and Diagnostic Accuracy} 

\author{
\IEEEauthorblockN{Garvit Goel
%\textsuperscript{\textsection}
}
\IEEEauthorblockA{\textit{Dept. of ECE} \\
\textit{Virginia Tech} \\
garvit217@vt.edu}
\and
\IEEEauthorblockN{Jingyuan Qi
%\textsuperscript{\textsection}
}
\IEEEauthorblockA{\textit{Dept. of CS} \\
\textit{Virginia Tech} \\ 
jingyq1@vt.edu}
\and
\IEEEauthorblockN{Wu-chun Feng}
\IEEEauthorblockA{\textit{Dept. of CS, ECE, and BEAM} \\
\textit{Virginia Tech} \\ 
wfeng@vt.edu}
\and
\IEEEauthorblockN{Guohua Cao}
\IEEEauthorblockA{\textit{Dept. of CS} \\
\textit{Virginia Tech} \\ 
ghcao@vt.edu}
}
%\author[\firstAuthorLast]{\Authors} %This field will be automatically populated
%\address{} %This field will be automatically populated
%\correspondance{} %This field will be automatically populated
%\extraAuth{}% If there are more than 1 corresponding author, comment this line and uncomment the next one.
%\extraAuth{Wu-chun Feng, wfeng@vt.edu}

\maketitle

\begin{abstract}
We present a deep-learning based computing framework for fast-and-accurate CT (DL-FACT) testing of COVID-19. Our CT-based DL framework was developed to improve the testing speed and accuracy of COVID-19 (plus its variants) via a DL-based approach for CT image enhancement and classification. The image enhancement network is adapted from DDnet, short for DenseNet and Deconvolution based network. To demonstrate its speed and accuracy, we evaluated DL-FACT across several sources of COVID-19 CT images. Our results show that DL-FACT can significantly shorten the turnaround time
from days to minutes and improve the COVID-19 testing accuracy up to 91\%. DL-FACT could be used as a software tool for medical professionals in diagnosing and monitoring COVID-19.

%Compared to the conventional reverse-transcription polymerase chain reaction (RT-PCR) test, DL-FACT can provide higher diagnostic accuracy and faster turnout time in detection of COVID-19.
% The COVID-19 pandemic has highlighted the importance of testing and monitoring as early and accurately as possible. However, the RT-PCR test has protracted turnaround time that could take days and compromised test accuracy that is as low as 67\%. 

%\tiny
% \keyFont{\section{Keywords:} 

%All article types: you may provide up to 8 keywords; at least 5 are mandatory.

\end{abstract}

%%%%%% Main Text %%%%%%

\section{Introduction}

Since the discovery of COVID-19 in December 2019, the world has recorded more than 200 million confirmed cases and nearly 4 million deaths of COVID-19 \citep{Dong2020}\citep{covid19-map}. Moreover, because roughly half of those infected people do not show symptoms (i.e. asymptomatic subjects) and act as unwitting transmitters of the virus  \citep{plater2020asymptomatic}, the number of unrecorded cases and the total number of infections are unknown. A study from the US Centers for Disease Control and Prevention (CDC) shows that 59\% of COVID-19’s transmission are asymptomatic  \citep{Johansson2021}. 
Furthermore, the accuracy of the widely used 2019 COVID-19 testing technique, real time polymerase chain reaction (RT-PCR) test, can be as low as 67\%, meaning one-third of the results could be false negative \citep{kucirka2020variation}. Additionally, new strains of the virus - such as the Delta and Omicron Variants - are still fast spreading. Rapid and accurate diagnostic tests for COVID-19 are very necessary to contain the spread of the COVID-19 disease.

% Background in biomedicine
Lung CT scans of COVID-19 patients revealed a number of prevalent features (also known as radiologic or CT abnormalities), such as ground-glass opacity (GGOs), linear opacity, consolidations, signs of reversal halo, and crazy-paving patterns  \citep{Wei2020}. These lung CT features have been found to present in more than 70$\%$ of COVID-19 cases confirmed by RT-PCR using multinational datasets  \citep{Harmon2020}. Although these features appear similar to the symptoms associated with certain other lung diseases, there are still some features that have been uniquely identified with COVID-19. For example, GGOs are commonly seen in CT images of tuberculosis patients, but bilateral GGOs that are clustered in the anterior segment of the lung are typically associated with COVID-19 \citep{Zhou2020}. These abnormal lung CT features can be identified by healthcare professionals to determine whether a patient has COVID-19. It is well known that deep-learning (DL) based AI technology in medical imaging can help in image ehancement, feature extraction and recognition. Therefore, a well-developed DL model should be able to help CT imaging as well as classification of COVID-19 positive and negative patients.

% our method
Here, we propose a DL-based computing framework that could potentially enable DL-based fast-and-accurate CT (hereafter termed DL-FACT) imaging of COVID-19. DL-FACT employs a DL neural network called DD-net to enhance lung CT images in the image domain \citep{Zhang2018}, so that the hallmark radiological features associated with COVID-19 pneumonia become more eminent. Then, the enhanced lung CT images are fed into another DL-based neural network to automatically detect COVID-19 pneumonia \citep{Harmon2020}. The former  DL network is referred as image enhancement AI and the latter DL network is referred as image analysis AI. Together, they form the DL-FACT framework. The enhancement AI is implemented as a 2D version of the DD-net based network, known as 2D-DDnet, as well as a 3D version of the DD-net based network, known as 3D-DDnet. The analysis AI contains two sub-models: an image segmentation sub-model and an image classification sub-model. The segmentation sub-model is used to segment the lung region from a volumetric CT scan, and the classification sub-model is used to predict probability of COVID-19 pneumonia for each segmented lung. The enhancement AI and the analysis AI operate independently from each other for CT image enhancement and CT scan classification, respectively.

% structure
The rest of this paper is organized as follows. In section 2, we describe our method in details, including data collection and preprocessing, DL model architectures, their training as well as evaluations. In section 3, we compare the quality of the original lung CT images and the enhanced lung CT images that are improved by our enhancement AI, and then compare the COVID-19 classification accuracy based on original and enhanced images utilizing a previously validated image analysis AI. Summary and discussion of our work are presented in section 4.

\section{Materials and Methods}
% In this study, we presented two AI tools: enhancement AI for the reconstruction of high-quality images and analysis AI for diagnosis. Some clinical image data are used to train neural networks and evaluate AI performance. In order to evaluate the performance of the enhancement AI on diagnosis, we also designed an evaluation framework to quantify the enhancement effect of the enhancement AI on the classifier (analysis AI).
The DL-FACT framework employs two AI tools: enhancement AI that is used to enhance lung CT images, and analysis AI that is used to detect COVID-19. After a short description of the data source used for the developments of the AIs, we explain the DL-FACT framework in detail, including the training and testing strategies for each AI and its sub-models.
\subsection{Data}
% In order to train and test our network and models, we collected lung CT scans from healthy people and confirmed COVID-19 patients from various public databases. Furthermore, some low-quality CT scans were simulated based on real images for experimental comparison.
In order to train and test the AIs in the DL-FACT framework, we collected a total of 124,070 lung CT images from a variety of public data sources, of which 53,670 images are from COVID-19-positive patients and 69,800 images are from COVID-19-negative patients. In addition, because it is unrealistic to have a high-quality CT scan and a correspondingly low-quality CT scan from the same patient, corresponding low-quality CT images are numerically simulated for network training. All images were preprocessed to the same format and fed to our AI models for training and testing. The clinical CT images and their correspondingly low-quality CT image, as well as the image preprocessign details, are described as follows.
\subsubsection{Clinical CT Images}
% For training the enhancement model, CT scans from 4 public biomedical image data sources are corrected. Each source contains some 3D volume CT scans (with size 512 x 512 x N) of lung region. Four public biomedical image data sources are listed below:
To train the enhancement AI, lung CT scans of COVID-19-positive patients are collected from two different public biomedical databases: BIMCV-COVID19+ dataset from the Medical Imaging Databank of the Valencia Region, as well as the Medical Imaging and Data Resource Center (MIDRC) dataset from the Radiological Society of North America (RSNA). To help train the analysis AI and evaluate the effect of the enhancement AI on improving COVID-19 diagnostic accuracy, lung CT scans of COVID-19-negative patients are collected from another two public databases: Mayo Clinic database from the Mayo Clinic Research Core Facilities, and Lung Image Database Consortium (LIDC) database. Both the Mayo Clinic and LIDC databases were made available before December 2019 when the first COVID-19 cases were discovered. All the databases offer 3D volumetric lung CT scans with the size of each volume at 512 x 512 x N, where N is number of axial slices. The four public biomedical databases are further explained as follows:

\textbf{Mayo Clinic Data} : The Mayo Clinic data contain 2,886 full and quarter dose 512 × 512 CT images from 10 patients \citep{mayo_clinic}. The quarter dose images were reconstructed from the projection data after Poisson noise was inserted into the full dose projection data for each full dose case. 

\textbf{BIMCV-COVID19+} The BIMCV-COVID19+ dataset is a large dataset that contains lung X-ray images and CT images of COVID-19 patients. From the BIMCV-COVID19+ and MIDRC datasets, we selected 44,384 COVID-19 CT slice images of 222 patients in NII format.

\textbf{MIDRC} The MIDRC dataset \citep{midrc} is hosted by RSNA. RSNA is collaborating with radiology organizations around the world to build a repository of COVID-19 imaging data for research. From the BIMCV-COVID19+ and MIDRC datasets, we selected 7000 COVID-19 CT slice images of 35 patients in NII format.

\textbf{LIDC-IDRI}: The LIDC dataset consists of diagnostic and lung cancer screening CT scans. From the LIDC dataset, we selected 69,800 CT slice images of 349 patients in DICOM format. 

\subsubsection{Simulation of Low-quality CT Images}

    In reality, most publicly available CT image datasets are of high quality, but it is easy for one to agree that low-quality CT images would be generated in real practice (e.g. low-dose CT images with higher noises). Therefore, there is a need to improve those low-quality CT images, for better diagnostic accuracy. This is exactly our motivation in developing our CT image enhancement AI. To train and test the enhancement AI, we simulated the low-quality CT scanning process by modeling those important imaging physics and generated a large number of simulated CT images at low quality.
    
    The details of the simulation process are as follows. From each original high-quality CT image, we generate the corresponding projection data using Beer's law and Siddon's ray-driven forward projection method \citep{siddon1985fast}. The X-ray source energy is assumed to be monochromatic at 60 keV. Poisson noise was simulated according to the formula $P_i = Poisson\{ b_i*e^{-l_i}\}$, i=1, 2, ..., N, where $P_i$ is the detector measurement along the $i^{th}$ ray path, $b_i$ the blank scan factor, and $l_i$ the line integral of attenuation coefficients along the $i_{th}$ ray Path. In this experiment, we assume that there is no electronic noise from the detector readout. The level of Poisson noise can be adjusted by setting the number of photons per ray for the blank scan factor $b_i$. In this study, we used $b_i=10^6$ photons per ray. The distance between the x-ray source and the detector is 1500 mm, and the distance between the x-ray source and the center of the object is 1000 mm. The detector1,024 pixels for X-ray detection. A total of 720 projections were simulated over a 360-degree circular scan trajectory using 0.5-degree step angle. After the projection data are simulated, filtered back projection (FBP) is used to reconstruct low x-ray dose CT images from the simulated projection data.

\subsubsection{Image Preprocessing}
Images from the public databases are in DICOM or NII formats, which are 3D image formats. However, our enhancement AI works only with 2D slice images in TIFF format as input, so format conversion is required for training and testing the enhancement AI. Conversion operation is implemented with software ImageJ. For DICOM format files, converting DICOM to TIFF is done by the “Batch + Convert” function of ImageJ. For NII format files, converting one 3D NII file to a series of 2D TIFF files is done using the “Save as + Image Sequence” function of ImageJ. Notice that both DICOM and NII files are stored with the datatype of 16-bit unsigned, so this datatype needs to be defined during the conversion.

\begin{figure}[H]
    \centering
    \includegraphics[width=0.45\textwidth]{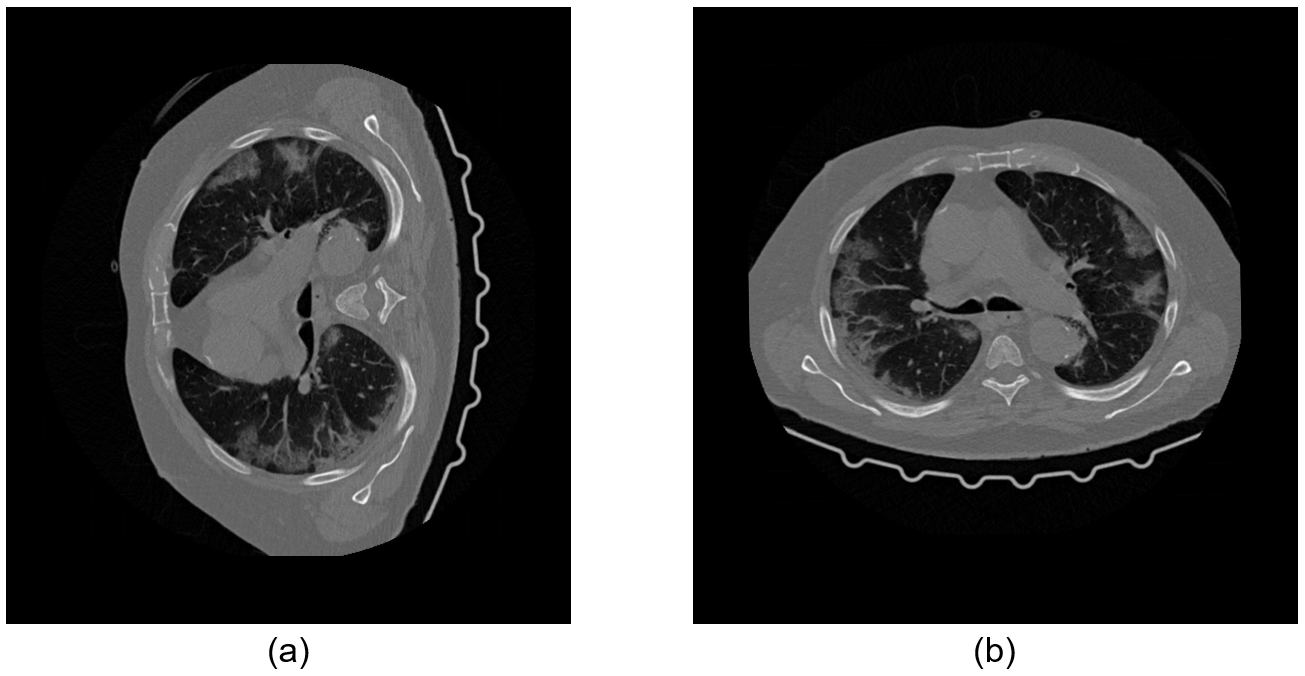}
    \caption{Orientation alignment of CT images during image preprocessing. \textbf{(a)} original CT image. \textbf{(b)} re-aligned CT image.}
    \label{orientation}
\end{figure}

Orientations of the CT images from different data sources are different. In the lung region, symptoms caused by COVID-19 are more commonly found near the back and on both sides of the lung. To avoid classification errors caused by different orientations, the orientations of all the images are re-aligned uniformly. In the axial view, the back of the chest is at the bottom of the re-aligned image, and the left lung is in the re-aligned image's right. Atypical re-aligned COVID-19 CT image is shown in Figure \ref{orientation}.

\subsection{DL-FACT Framework}
\subsubsection{Enhancement AI}
\label{se:enhancement_gen}

\begin{figure}[!htb]
 \centering
 \includegraphics[width=0.45\textwidth]{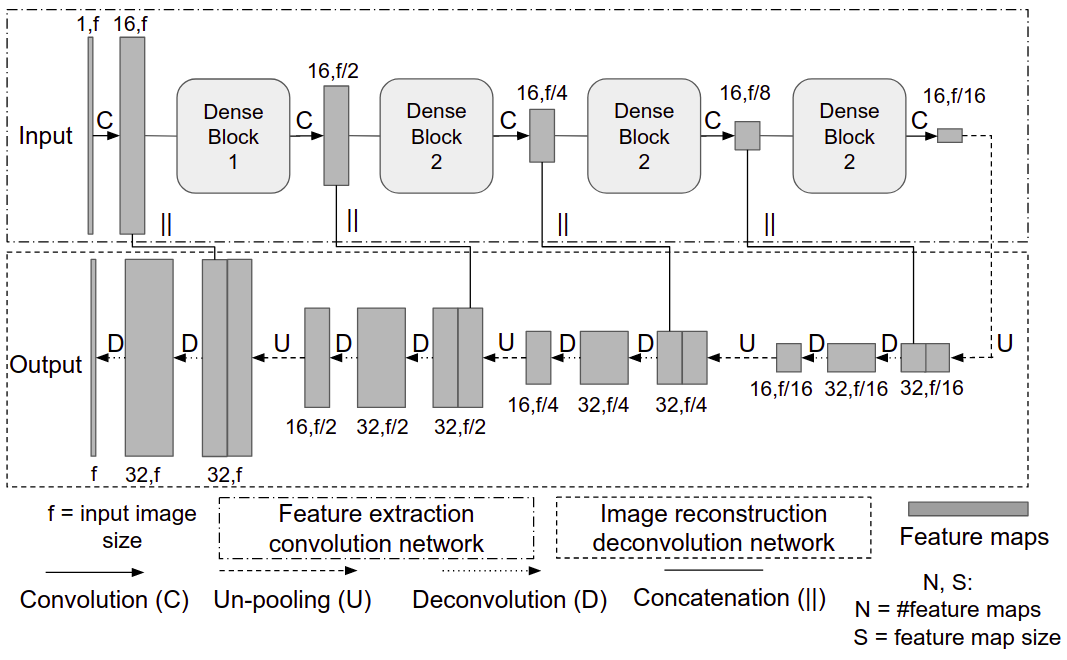}
 \caption{The architecture of DDnet }
 \label{fig:dd_net_arch}
 \vspace{-9pt}
 \end{figure}

The image enhancement AI is based on a convolution neural network called DD-Net (short for DenseNet and Deconvolution combined network)~ \citep{Zhang2018}. The network uses 4 dense blocks~ \citep{huang2017densely} for feature extraction and 8 deconvolution layers for image reconstruction from the extracted features. The architecture of the network is shown in Figure~\ref{fig:dd_net_arch}. In this project, we implemented both 2D and 3D versions of the DD-Net. The 2D version of the enhancement AI requires less computing resources and takes less time for training and inference; The 3D version is supposed to leverage the spatial correlation among the consecutive 2D CT slices of a volumetric 3D CT scan and extract features in 3D space.

\textbf{2D-DDnet}
\label{sec:enhancement2D}

2D DDnet network inputs $512\times512$ CT images and outputs enhanced images of the same size. Both convolution and deconvolution layers in the network use $5\times5$ filters with zero paddings and a stride of 2. Pooling operations use $3\times3$ filters with a stride of 2. Filters were initialized with a random Gaussian distribution with a mean of zero and a standard deviation of 0.01.

The loss function used for back propagation during the training is a composite function that combines mean square error (MSE) and multi-scale structural similarity index metric (MS-SSIM). The MS-SSIM~\cite{wang2004} compares the luminance, contrast, and structural similarity between two images. The loss function, $L$, is given by Equation~\ref{eq:loss}.

\begin{equation}
    \label{eq:loss}
%    \scalebox{0.9}{\parbox{\linewidth}
%    \scalebox{0.90}{
    \mathscr{L} = 
        ||y-f(x)||_2^2  + 0.1  \times (1 - L_{MS-SSIM}(Y, f(X)))
%    }
\end{equation}
where ${||y  - f(x)||_2^2}$ is the MSE term and $L_{MS-SSIM}$ is the MS-SSIM term.

Network weights are updated using the ADAM optimizer~ \citep{Kingma2014}. The hyper-parameters are tuned by perturbing one parameter while keeping others fixed and analyzing the quantitative results. Optimal values of the hyper-parameters are shown in Table~\ref{tab:train_param2d}.

\textbf{3D-DDnet}
\label{se:3d_vol_enhance}

The CT image enhancement using a 2D convolution neural network~ \citep{Wurfl2018}\citep{cheng2017accelerated} ignores the correlation between consecutive 2D image slices in a volumetric 3D scan.. 3D convolution neural networks use 3D cognition to extract and analyze features present in 3D volumes. To that end, we implement 3D volumetric enhancement for CT scans using a 3D adaptation of the 2D DDnet. The architecture of the 3D-DDnet is similar to 2D-DDnet, except each layer is modified to process 3D (in some cases 4D) data.

For the network training, 3D-DDnet uses the same loss function as 2D-DDnet, defined in Equation~\ref{eq:loss}. For calculating MS-SSIM for 3D volumes, luminance, contrast, and structural similarity are calculated using a moving 3D window~ \citep{Zeng2012}, in contrast to 2D MS-SSIM, which uses 2D moving window.

Network weights are updated using the ADAM optimizer~ \citep{Kingma2014}. All filters are initialized with a random Gaussian distribution with a mean of zero and a standard deviation of 0.01. The hyper-parameters are tuned by perturbing one parameter while keeping others fixed and analyzing the quantitative results. Optimal values of the hyper-parameters are shown in Table~\ref{tab:train_param3d}.

3D-DDnet has about 5 million training parameters, i.e., $5\times$ the number of training parameters in 2D-DDnet. A large number of training parameters makes the network very sensitive to inputs and noise. This leads to an unstable network resulting in exploding and vanishing gradients. We solve this problem by using a batch size of 2 or more CT scans as input to the network. Batch normalization used in the network normalizes the inputs in each batch so that the network is less sensitive to the noise in input scans~ \citep{ioffe2015batch}.

\textbf{Training Strategies}

The computational requirement of 3D-DDnet restricts the size of 3D CT scans that the network can take as inputs. Enhancement of $2^{22}$ pixels sized CT scan using 3D-DDnet requires about 8 GBs of memory. Keeping this in mind, we experimented with various training strategies with different sizes of CT scans for training 3D-DDnet. CT scans of size $512\times512\times64$ are divided into four sub-volumes of size  $256\times256\times64$ each. These sub-volumes are used as four independent inputs for training the network. In order to remove the artifacts generated due to zero padding in convolution and deconvolution operations, these sub-volumes are padded in both x and y directions, with 16 additional pixels from original CT scan data.

To maintain a constant number of image slices per input CT scan, image slices in input CT scans are homogeneously selected at equal distances from each other. For example, for a CT scan containing 512 image slices, the input CT scan of 64 image slices consists of every $8^{th}$ image slice from the original CT scan.

\subsubsection{Analysis AI}
The analysis AI model consists of two sub-models: a segmentation sub-model and a classification sub-model. The segmentation sub-model is used to segment the lung region from the CT scans, and the classification sub-model is used to predict the possibility of COVID-19 for each CT scan. 

\textbf{Segmentation Sub-model}

% An anisotropic hybrid network (AHNet) adapted for 3D CT image segmentation \citep{Liu2017} is chosen as the segmentation sub-model. This model accepts 3D volume images as input and generates same-size 3D volume images in binary values (0 and 1). The region whose pixel values = 0 in the output of segmentation AI represents the lung region.
The segmentation sub-model is adapted from the 3D anisotropic hybrid network (AH-Net) \citep{Liu2018}, which is based on ResNet and adapted for 3D image segmentation tasks. AH-Net incorporates a feature decoder composed of anisotropic convolutional blocks. By connecting the dense connection layer in ResNet with the anisotropic convolution blocks in the feature decoder, Ah-Net can transfer convolutional features learned from 2D images to 3D anisotropic volume. The segmentation sub-model accepts 3D volume images as input and outputs binary-value images of the same shape. The areas with pixel values of zero in the output images represent the lung region in the inputs.

\textbf{Classification Sub-model}

The classification sub-model is a 3D image recognition network. Although 2D image recognition network has been extensively studied, it does not take account of the fact that in 3D CT scans one slice image should be related to its neighboring slices (i.e. image consistency). For example, the appearance or disappearance of COVID-19 features in the lung region should not be sudden but should be a continuous process. In a 3D classification network, due to the consistency of the images, if an infection slice suddenly appears in continuous slices and is recognized as positive, it is more likely a misrecognition and will be ignored.

In this work, the classification sub-model is adapted from DenseNet-121 network and is implemented for 3D volume classification in the nVidia Clara framework \citep{clara}. The 3D implementation of DenseNet-121 accepts 3D CT images of size 512×512×$N$ as input, where $N$ is the number of 2D image slices in one 3D CT scan. The input image needs to be in Hounsfield units. The transformation DenseNet-121 from 2D to 3D is an extension of the dimension in the input size, kernel size, and output size; For the convolution layer, the extension is simply added on dimension at the z-axis. For example, the kernel size is from 3 x 3 in 2D to 3 x 3 x 3 in 3D. Similar to the convolutional layer, the extension of the pooling layer is adding a dimension to the filter.

\textbf{Training Strategies}

Because it is difficult to obtain a large number of labeled COVID-19 CT images and delineated lung regions from lung CT scans, in this work we used the pre-training analysis AI model described by Harmon et al. \citep{Harmon2020}. The implementation of this analysis AI is made available by Nvidia \citep{clara}. To fine-tune this pre-trained analysis AI with our data, the analysis AI is re-trained with 317 volumetric CT scans in shape of 512 x 512 x N (N is the number of 2D slice images in one 3D volume), consisting of 145 true-positive cases and 172 true-negative cases. The training parameter values are listed in Table \ref{tab:cla_train_para}. 

\subsubsection{Framework Integration}
After the training of each network described above, all the AIs were integrated into the DL-FACT framework. In an ideal scenario, the DL-FACT framework takes a low-quality CT scan as the input, enhances the scan with the enhancement AI, and then the output from the enhancement AI is fed into the analysis AI for COVID-19 classification. The classification AI requires the segmented lung region from the CT scan, which can be obtained from the CT images before or after the enhancement. Here, we chose to produce the segmented lung region from the CT images before the enhancement.

\subsection{Evaluation Method}

% To test the improvement in COVID-19 diagnostic accuracy from the proposed DL-FACT framework, we compared the COVID-19 diagnostic accuracy of a testing dataset, before and after the image enhancement by the enhancement AI. The testing dataset consists of 36 CT scans of COVID-19-positive patients and 59 CT scans of healthy people. Figure \ref{fig:testframe} is the overall flow chart for the testing strategy.
In order to evaluate the effect of enhancement AI, we compared the quality of the images before and after the enhancement, both qualitatively and quantitatively. Additionally, we tested the effect of enhancement AI on diagnostic accuracy.

\subsubsection{Image Quality Improvement}
To evaluate the image quality itself, we used mean square error (MSE) and Multi-Scale Structural Similarity Index Metric(MS-SSIM). The two metrics are defined as follows. 

\textbf{MSE}

For image A and B, take the square of the difference between each pixel value in image A and the corresponding pixel in image B, sum up these squares of difference and divide the sum by the total number of pixels.\citep{Lehmann1998}

\textbf{MS-SSIM}

MS-SSIM compares the luminance, contrast, and structure similarity between two images. Its formula is as below:
    \begin{equation}
        SSIM(x,y) = \frac{(2\mu_x \mu_y +c_1)(2\theta_{xy}+c_2)}{(\mu_x^2 + \mu_y^2 +c_1)(\theta_x^2 + \theta_y^2 + c)2)}
    \end{equation}
where $\mu_x$ is the average of x, $\mu_y$ is the average of y, $\theta_x^2$ is the variance of x, $\theta_y^2$ is the variance of y, $\theta_{xy}$ is the covariance of x and y. $c_1 = (k_1L)^2$ and $c_2 = (k_2L)^2$ are two variables to stabilize the division with weak denominator, and $L$ is the dynamic range of the pixel-values \citep{wang2004}.

\subsubsection{Improvement on COVID-19 diagnostic accuracy}
To evaluate the effect of enhancement AI on COVID-19 diagnostic accuracy, the original images and their correspondingly enhanced images are both fed to the same well-trained analysis AI. Figure 3 shows the overall flow chart of the evaluation strategy. The testing dataset consists of 177 lung CT scans of negative patients and 93 lung CT scans of positive patients.

% \textbf{Diagnostic accuracy of the original image}

% Firstly, the original images are segmented by the segmentation sub-model and generate the segmentation masks of the original images. The original images and the segmentation masks are received by the classification sub-model as input values. The classification sub-model will output whether the input CT scans come from COVID-19 patients. Comparing the output from the classifier with ground-truth, the diagnostic accuracy based on original images could be calculated.
First, the segmentation sub-model was used to segment and generate the segmentation masks from the original CT images. Then, the classification sub-model accepts a CT scan and its corresponding segmentation mask as inputs, and outputs the binary result of the input CT scan that indicates a probability score for the CT scan. We run the same classification sub-model on both the original CT images and the correspondingly enhanced CT images. The classification results are compared to the ground truth labels, which gives us the improvement on COVID-19 diagnostic accuracy due to our enhancement AI.

\textbf{Diagnostic Accuracy}

To quantitatively characterize the improvement on COVID-19 diagnostic accuracy, we calculated the following diagnostic metrics:
%     \textbf{True positive (TP):}
%       predicted as positive, while actual class is positive;
%     \textbf{True negative (TN):}
%       predicted as negative, while actual class is negative;
%     \textbf{False positive (FP):}
%       predicted as positive, while actual class is negative;
%     \textbf{False negative(FN):}
%       predicted as negative, while actual class is positive;
% Based on these four values, we define following evaluation values:
    
    \textbf{Precision:} ratio of correctly predicted positive to the total predicted positive samples.
    
    \textbf{Recall:} known as true positive rate, the ratio of correctly predicted positive to the all testing samples in actual positive class.
    
    \textbf{Accuracy:}ratio of correctly predicted observation to the total testing samples.
    
    \textbf{F1-score:} weighted average of Precision and Recall; more fair with an uneven class distribution.
      
      \begin{equation*}
      \begin{aligned}[l]
          Pre &= \frac{TP}{TP+FP}\\
          Rec &= \frac{TP}{TP+FN}\\
          Acc &= \frac{TP+TN}{TP+FP+FN+TN}\\
          F1 &= 2\cdot \frac{Rec \cdot Pre}{Rec + Pre}
          \label{equ:statis}
      \end{aligned}
      \end{equation*}
      
where TP is true positive, TN is true negative, FP is false positive and FN is false negative. Furthermore, to evaluate the comprehensive performance of classification, we generated a ROC curve and calculated the AUC value for the classifier. 
    \textbf{ROC curve }
      is the receiver operating characteristic curve, a graphical plot that illustrates the diagnostic ability of a binary classifier system as its discrimination threshold is varied.
    \textbf{AUC value }
    is the area under the ROC curve, equal to the probability that a classifier will rank a randomly chosen positive instance higher than a randomly chosen negative one. The higher AUC value, the better classifier.
The package \emph{sklearn} is used to generate the ROC curve and AUC values.

Comparing the difference of precision, recall, accuracy, f1-score, and AUC value from results based on original and enhanced images, we can evaluate the effect of enhancement AI on COVID-19 diagnosis.

\section{Results}

\subsection{Image Quality Improvement from Enhancement AI}

\textbf{2D-DDnet}

%For evaluating the ability of the \textsf{Enhancement AI} in improving the quality of CT images, we quantify the network accuracy using the MSE and MS-SSIM between the original CT image and enhanced CT image.

Figure~\ref{fig:2D_enhancement}(a) shows the result of enhancing lung CT images from the Mayo Clinic lung CT data. CT images in Mayo Clinic dataset are grouped into pairs of high-dose (i.e. high-quality) CT images and low-dose (i.e. low-quality) CT images. We can observe that the enhancement AI removes some noise present in low-dose CT images while retaining fine details. Figure~\ref{fig:2D_enhancement}(b) shows the results of enhancing CT images from a simulated data set. \textsf{Enhancement AI} removes the streaking artifacts and noise present in the image, which was FBP-reconstructed from the simulated projection data at non-optimum acquisition conditions (i.e. sparse view and low dose). Qualitatively, the enhanced images for both data sets have well-defined boundaries and fine details.

Quantitatively, the \textsf{Enhancement AI} achieved an average of 98.7\% multi-scale structural similarity between the high-quality target image and enhanced image for CT images in the testing data set. Table~\ref{tab:ddnet_quantitative} summarizes the quantitative results of the \textsf{Enhancement AI}.

To further illustrate the effectiveness of CT image enhancement, Figure~\ref{fig:diff_maps} shows the absolute difference maps between the low-quality image and high-quality image, and enhanced image and high-quality image for two data sets. It is apparent that the enhancement AI effectively removes the missing projection artifact and noise present in low-quality CT images. 

% \begin{figure}[htb]
%   \centering
%   \includegraphics[width=0.9\textwidth]{figures/diff_maps2.png}
%   \caption{Absolute difference maps for \textbf{(a)} Mayo clinic data set, and \textbf{(b)} Simulated data set.}
%   \label{fig:diff_maps}
% \end{figure}

\textbf{3D-DDnet}

Similar to 2D-DDnet, we quantify the accuracy of 3D enhancement using the mean square error (MSE) and multi-scale structural similarity index metric (MS-SSIM) between the original CT images and the enhanced CT images. 

Results of 3D enhancement using one image slice from the CT scan of size $256\times256\times64$ pixels (originally $512\times512\times64$ pixels in size and resized to $256\times256\times64$ pixels), $512\times512\times16$ pixels, and $512\times512\times64$ pixels (input to the network as 4 separate tiles of size $256\times256\times64$) are shown in Figure~\ref{fig:3D_enhancement}. The \textsf{Enhancement AI} reconstructs high quality CT images by removing noise present in the input low-quality CT scan while retaining finer details. The removal of noise and artifacts from the low-dose CT scan is clearly visible in absolute difference maps, shown in Figure~\ref{fig:3d_diff}.

% \begin{figure}[]
%   \centering
%   \includegraphics[width=0.9\linewidth]{figures/3d_scan_sum.png}
% \caption{\textbf{(a)} 3D volumetric of size 256x256x64 enhancement using 3D-DDnet. \textbf{(b)} 3D volumetric of size 512x512x16 enhancement using 3D-DDnet. \textbf{(c)} 3D volumetric of size 512x512x64 enhancement using 3D-DDnet.}
% \label{fig:3D_enhancement}
% \end{figure}

% \begin{figure}[!t]
%  \centering
%  \includegraphics[width=0.60\linewidth]{figures/3d_diff3.png}
%  \caption{Absolute difference maps for CT image slice in \textbf{(a)} $256\times256\times64$ pixel sized CT scans, \textbf{(b)} $512\times512\times16$ pixel sized CT scans,  and \textbf{(c)} $512\times512\times512$ pixel sized CT scans.}
%  \label{fig:3d_diff}
%  \end{figure}

Quantitatively, 3D image \textsf{Enhancement AI} reduces MSE and improves MS-SSIM between high-quality target CT scan and enhanced CT scan, as shown in Table~\ref{tab:3d_enhancement_quant}. 3D-DDnet achieves the highest accuracy. 

%This is because of the following reasons. 

%with the dataset containing CT scans of size $256\times256\times64$ pixels, as compared to CT scans of size $512\times512\times16$ pixels and $512\times512\times64$ pixels

% \input{table/3d_enhancement_quant}

%MS-SSIM between high-quality target and low-quality input CT scans of size $256\times256\times64$ pixels is higher than that of $512\times512\times16$ and $512\times512\times64$ pixel-sized CT scans. 

%\begin{itemize}
%    \item Convolution network extracts finer features with higher quality CT scan as input for volumetric reconstruction via deconvolution network.
%    \item Dataset containing CT scans of size $256\times256\times16$ pixels are more isotropic in terms of scan size, as compared to the dataset containing CT scans of size $512\times512\times16$ pixels.
%    \item Tiling of CT scans, of size $512\times512\times64$ pixels, generates more variability in input than inputs containing whole CT scans. The variability in input requires more complex functional mapping from input to output. Since the architecture of the network is fixed, simpler inputs achieve higher accuracy with the trained network.
    
%\end{itemize}

% \input{table/comp_2d_3d_ddnet}

\textbf{Comparison of 2D Image and 3D Volumetric Enhancement}

The quantitative comparison of CT scan enhancement using 2D and 3D-DDnet, trained and tested on the same dataset, is shown in Table~\ref{tab:comp_2d_3d_ddnet}. 3D-DDent outperforms 2D-DDnet in terms of accuracy, i.e., in reducing the MSE and improving MS-SSIM between. This is because, in addition to enhancing 2D features in CT image slices, 3D-DDnet enhances $3^{rd}$ dimensional correlation features between the image slices present in the CT scans, which 2D DDnet ignores. The enhancement of features in y-z plane, by trained networks, is shown in Figure~\ref{fig:enhancement_z}. 3D-DDnet enhances the features along the z-direction, while there is no enhancement along the z-direction by 2D-DDnet.

\subsection{Diagnostic Accuracy Improvement from Analysis AI}
Figure \ref{fig:2} presents the accuracy plot and ROC curve. Qualitatively, comparing the ROC curve based on original and enhanced images, we can see that there is a larger area under the ROC curve of results based on enhanced images. The threshold value = 0.061 and 0.026 are the optimal thresholds when the classifier uses the original image and the enhanced image, respectively, making the accuracy of the classification result highest. %When the threshold is set to the optimal value, the precision, recall, accuracy, and f1-score of the classification results based on the enhanced image are higher than those based on the original image.

Quantitatively, after enhancement, the true-positive cases number increases from 28 to 31, the true-negative number increases from 54 to 55. The average positive probability score (APPS) of positive ground-truth cases increases from 0.4364 to 0.5500 (26$\%$), while the APPS of negative ground-truths stay almost the same value. All of the accuracy, precision, recall and AUC values are increased in the enhanced case. %The area under the enhanced ROC curve (AUC) increased from 0.8901 to 0.9419, the accuracy from 0.8485 to 0.8857, the recall from 0.7778 to 0.8611, the f1-score from 0.8116 to 0.8732, and the accuracy from 0.8632 to 0.9053. 
Detailed results are listed in Table \ref{tab:statResult} .

As can be seen from the above statistics, our image enhancement AI improves the classification results of the classifier on both accuracy, precision, and recall, where the increase of recall is much larger. %The accuracy increased by 4.38 $\%$ , indicating a reduction in the number of misclassifications in the test samples identified as positive, and the recall increased by 10.71 $\%$, indicating that the classifier is more likely to classify positive cases from all true-positive cases. 
% There are two potential reasons: 1. The change of FN is much larger than it of FP. Figure \ref{fig:pps} shows the positive probability scores of both positive and negative case samples. %The left two columns show the samples from enhanced images, and the right two columns show the samples from original images. 
Because most incorrect classifications happen as a false negative, which means the classifier recognizes the actually positive sample as negative. On the other hand, most actual negative samples are recognized correctly. After enhancement, the improvement of actual positive recognition is more than that of negative. The decrease of false-negative numbers is larger than that of false-positive numbers. 
%According to Equation \ref{equ:statis}, less false negative FP change will lead to a less increasing of precision; 2. Because most incorrect classifications happen as a false negative, so the recall in the original case is relatively smaller, which will easier to cases a relatively larger improvement.

Both accuracy and F1-score are improved. These two comprehensive index values indicate that the Enhancement AI improved the performance of the classifier comprehensively. 

% \begin{table}[h]
% \centering
% \caption{Statistical Result of Testing Framework}
% \label{tab:statResult}
%     \begin{tabular}{cccc}
%     \hline \hline
%                       & \textbf{Original CT Image} & \textbf{Enhance CT Image} & \textbf{Improvement} \\ \hline \hline
%     % \textbf{Positive APPS} & 0.4364                     & 0.5500                    & 26\%               \\
%     \textbf{Precision} & 0.8485                     & 0.8857                    & 4.38\%               \\
%     \textbf{Recall}    & 0.7778                     & 0.8611                    & 10.71\%              \\
%     \textbf{F1-score}  & 0.8116                     & 0.8732                    & 7.59\%               \\
%     \textbf{Accuracy}  & 0.8632                     & 0.9053                    & 4.88\%               \\
%     \textbf{AUC}       & 0.8901                     & 0.9419                    & 5.82\%               \\ \hline \hline
%     \end{tabular}
% \end{table}

% \begin{figure}[h]
%     \centering
%         \includegraphics[width=0.9\textwidth]{figures/roc_and_acc.png}
%         \caption{\textbf{(a)} Diagnostic Accuracy in different Thresholds. \textbf{(b)} ROC curve changing of classification sub-model between original CT image based and enhanced CT image based.}
%     \label{fig:2}
% \end{figure}

\section{Discussion}
In this work, we implemented a DL neural network (DD-net) to enhance the images of lung CT scanning, and trained an analysis AI which can recognize the lung CT scanning images of COVID-19 positive cases by supervised learning method. The DD-net was implemented in both 2D-DDnet and 3D-DDnet, the latter of which leverages the spatial correlation between 2D CT image slices in a 3D volumetric CT scan. The image enhancement AI is integrated with the image analysis AI to form an artificial intelligence framework called \emph{DL-FACT}.  After the image enhancement, the accuracy in detecting COVID-19 pneumonia from lung CT scans is significantly improved. In our tests using clinical lung CT scans, our image enhancement AI improved COVID-19 detection accuracy from 86$\%$ to 91$\%$, and 3D-DDnet based enhancement AI can further enhance the lung CT images.

% merits
Compared to the conventional RT-PCR detection, our DL-FACT framework can provide both higher overall accuracy and faster turnout time in diagnosing COVID-19. Conventional RT-PCR test requires testing materials, professional labor, as well as intensive logistics. This process inherently takes time. Our AI-based approach significantly reduces the required materials and labor, which greatly reduces the detection time (RT-PCR takes at least 4 hours and our detection framework takes about 5 minutes). More importantly, from sample collection to RT-PCR bench analysis, any errors in the many steps involved in this process could lead to false negative results. In contrast, our framework relies on a deep neural network called DD-net and an image analysis AI. DD-net is an image-domain-based image ehancement network. With careful preparation of training data that contain many labeled low-quality and high-quality lung CT image pairs, we can train DD-net to learn the correlation between low-quality lung CT images and high-quality lung CT images that are obtained from the same imaging scanner for the same patient. Then, during the testing phase, the well-trained DD-net can be used to enhance low-quality lung CT images into high-quality counterparts, making the details and features of the lung easier to recognize. With high-quality lung CT images, the confidence in identifying the hallmark features associated with COVID-19 pneumonia is improved, and hence improving the diagnostic accuracy of COVID-19 (RT-PCR has an overall sensitivity of ~67\% \citep{kucirka2020variation} and our detection framework has a diagnostic sensitivity of ~86.1\%.

A limitation of our DL framework is that the dataset of the truly negative images only comes from the LIDC-IDRI. The LIDC-IDRI consists of diagnostic and lung-cancer screening lung CT scans. This dataset was collected from 7 academic centers and 8 medical imaging companies over many years, and images in this dataset are uniformly screened and processed. However, the image data of all positive COVID-19 cases in this study are from clinical cases in different hospitals which have no uniform standard. Many external factors, such as the quality of the scanner performing the scan, the format of the data stored, and the image processing steps prior to sharing the data, will have a potential impact on the DL training and analysis. The goal of this study is to prove that the enhancement AI can improve image quality and have a beneficial effect on COVID-19 diagnosis. Since the data used for training and testing come from the same sources, we expect that these potential external factors will have a minimum effect on the evaluation of the enhancement step. Another general concern is that the data used in training our DL networks are not representative enough, which would lead to poor performance when being applied to practical clinical data. This concern is common for many state-of-the-art DL methods, and potential users of our DL framework need to use their real clinical data to retrain our DL model for practical use.

The COVID-19 pandemic has led to a constant demand for high-quality testing. The DL framework proposed in this paper could help improve the diagnostic accuracy of COVID-19 by improving image quality. Furthermore, the faster turnout time from our DL approach is also beneficial for containing this pandemic. 

\section*{Acknowledgments}
This research was supported in part by NSF IIS-2027607. We thank Dr. Cynthia McCollough, the Mayo Clinic, and the American Association of Physicists
in Medicine and grants EB017095 and EB017185 from the NIBIB, for providing the Mayo Clinic dataset used in this work. The authors acknowledge Advanced Research Computing at Virginia Tech for providing computational resources and technical support that have contributed to the results reported within this paper. URL: https://arc.vt.edu/.

\bibliographystyle{frontiersinSCNS_ENG_HUMS} % for Science, Engineering and Humanities and Social Sciences articles, for Humanities and Social Sciences articles please include page numbers in the in-text citations
\bibliography{test}

% \printbibliography

\section*{Figure captions}

 \begin{figure}[H]
    \centering
    \includegraphics[width=0.5\textwidth]{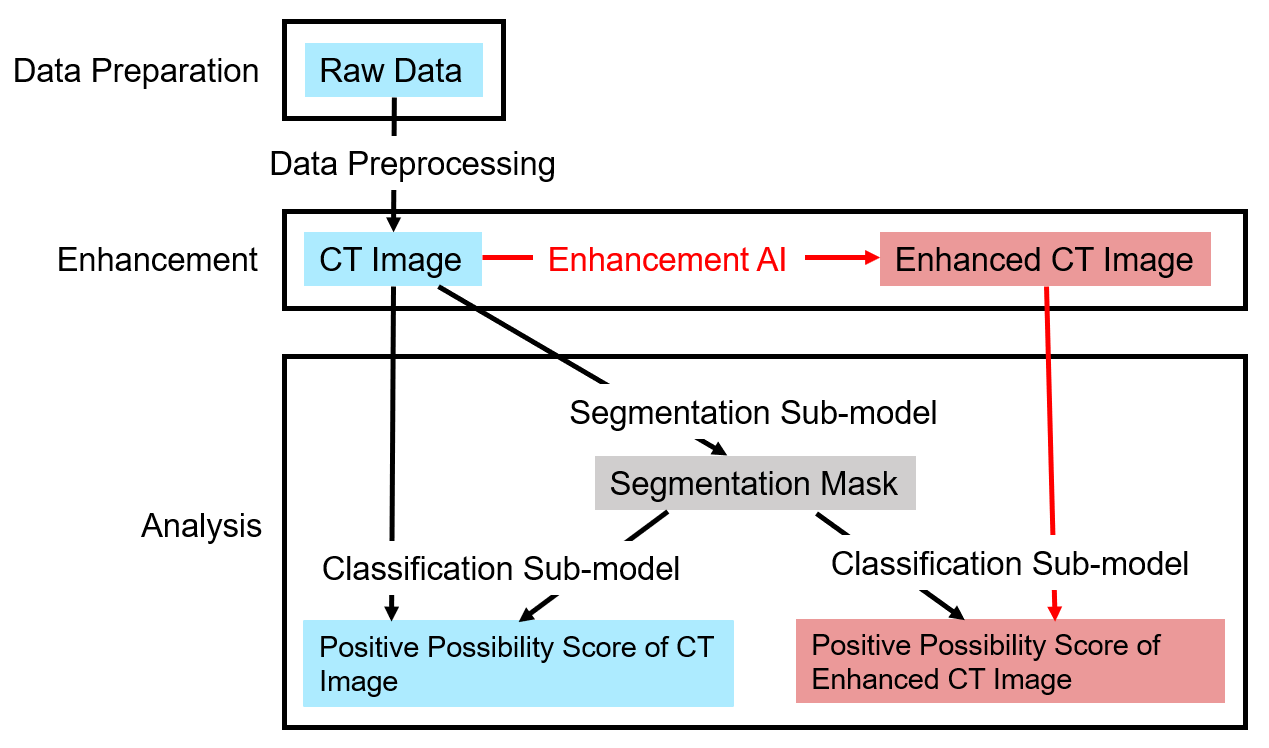}
    \caption{Overall flow chart for the testing strategy of the proposed DL-FACT framework. In the chart, analysis AI consists of segmentation sub-model and classification sub-model.}
    \label{fig:testframe}
\end{figure}
 
 \begin{figure}[H]
  \centering
  \includegraphics[width=0.95\linewidth]{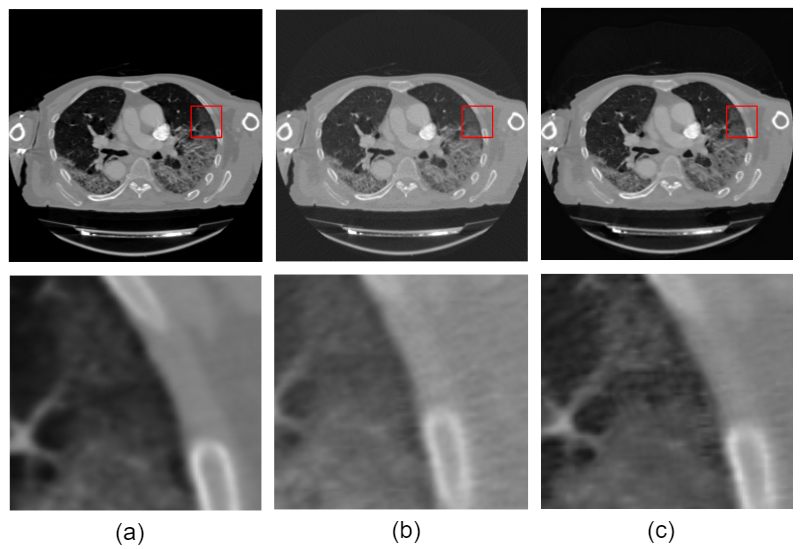}
  \caption{Global views and zoomed in views for the red box areas in the COVID CT images for (a) High quality CT image, (b) Low quality CT image, and (c) Enhanced images by 2D DD-net.}
  \label{fig:2D_enhancement}
\end{figure}
 
 \begin{figure}[H]
  \centering
  \includegraphics[width=0.5\textwidth]{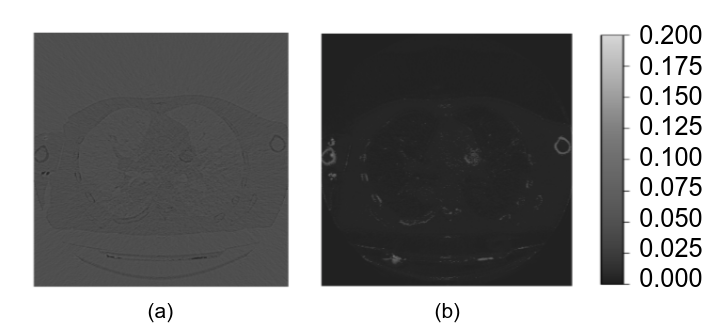}
  \caption{Absolute difference maps between (a) High quality CT image and low quality CT image, (b) High quality CT image and CT image enhanced by 2D-DDnet.}
  \label{fig:diff_maps}
\end{figure}

\begin{figure}[H]
  \centering
  \includegraphics[width=0.95\linewidth]{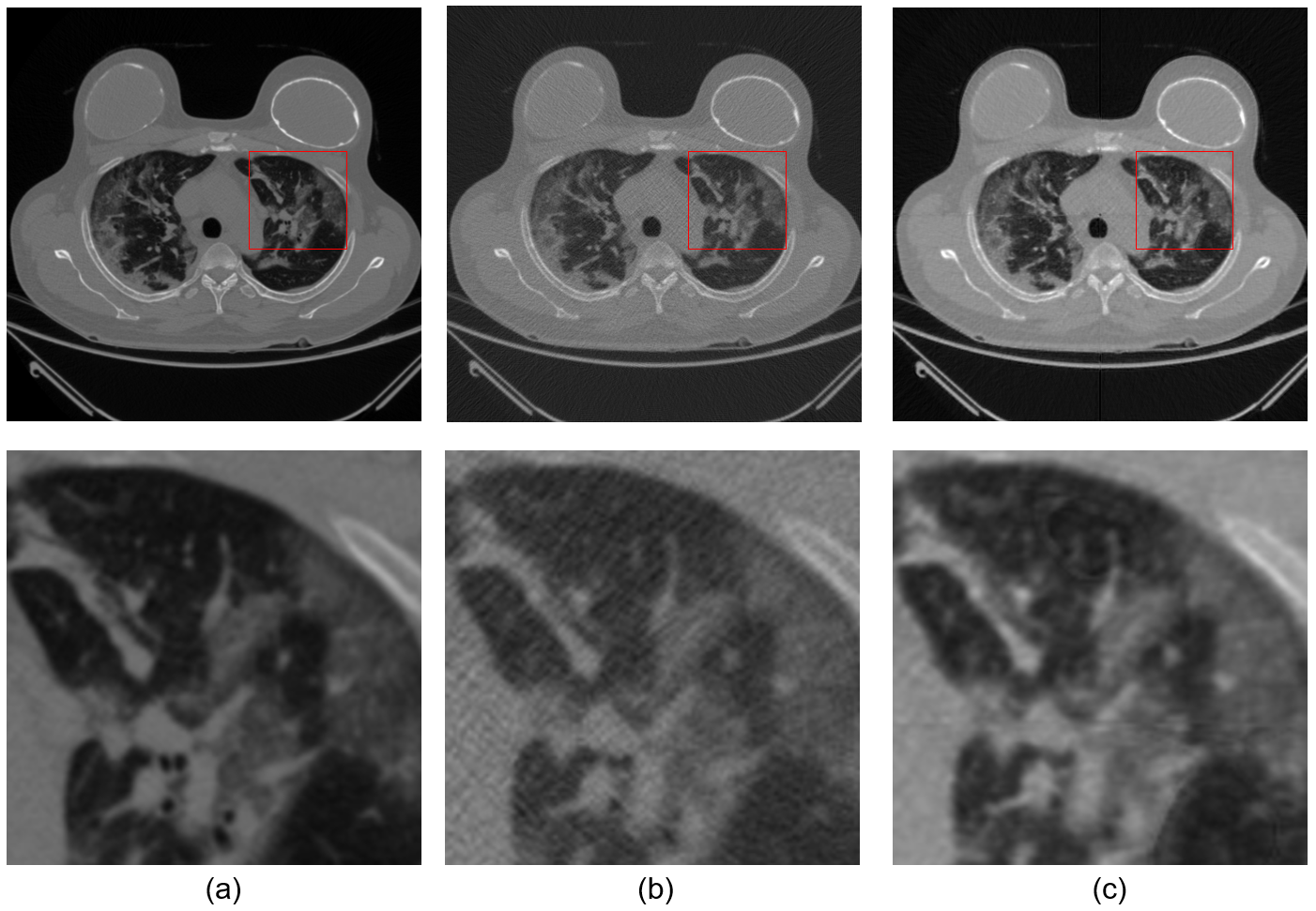}
\caption{Global views and zoomed in views for the red box areas in the COVID CT images for (a) High quality CT image, (b) Low quality CT image, and (c) Enhanced images by 3D DD-net.}
\label{fig:3D_enhancement}
\end{figure}

\begin{figure}[H]
 \centering
 \includegraphics[width=0.95\linewidth]{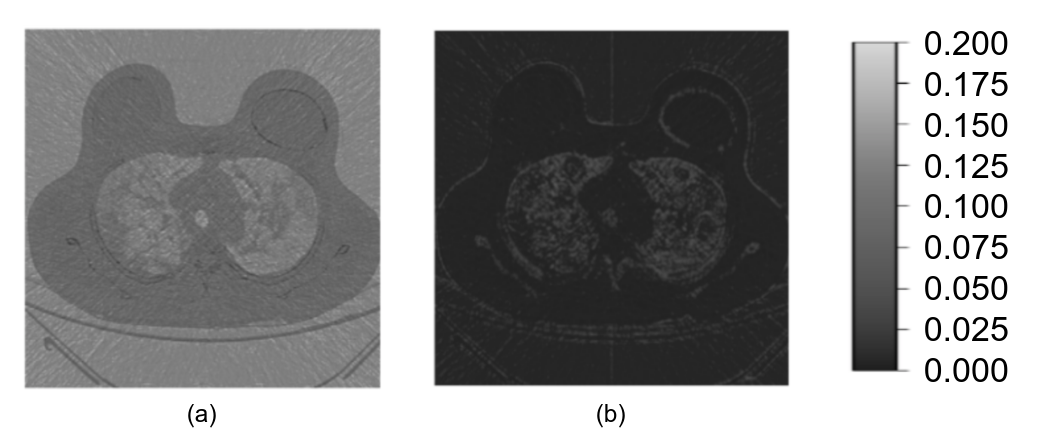}
 \caption{Absolute difference maps between (a) High quality CT image and low quality CT image, (b) High quality CT image and CT image enhanced by 3D-DDnet.}
 \label{fig:3d_diff}
 \end{figure}
 
 \begin{figure}[!h]
 \centering
 \includegraphics[width=0.4\linewidth]{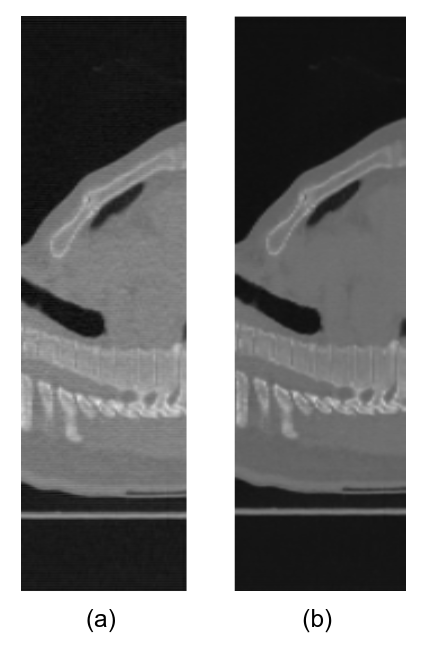}
 \caption{Enhancement of features in y-z plane by 2D and 3D-DDnet. CT Data source: MIDRC \citep{midrc}}
 \label{fig:enhancement_z}
 \end{figure}
 
% \begin{figure}[H]
%     \centering
%     \includegraphics[width=0.9\textwidth]{figures/appear.pdf}
%     \caption{Sample of 2D Slices Consistency in 3D Image}
%     \label{fig:appear}
% \end{figure}

\begin{figure}[H]
    \centering
        \includegraphics[width=0.5\textwidth]{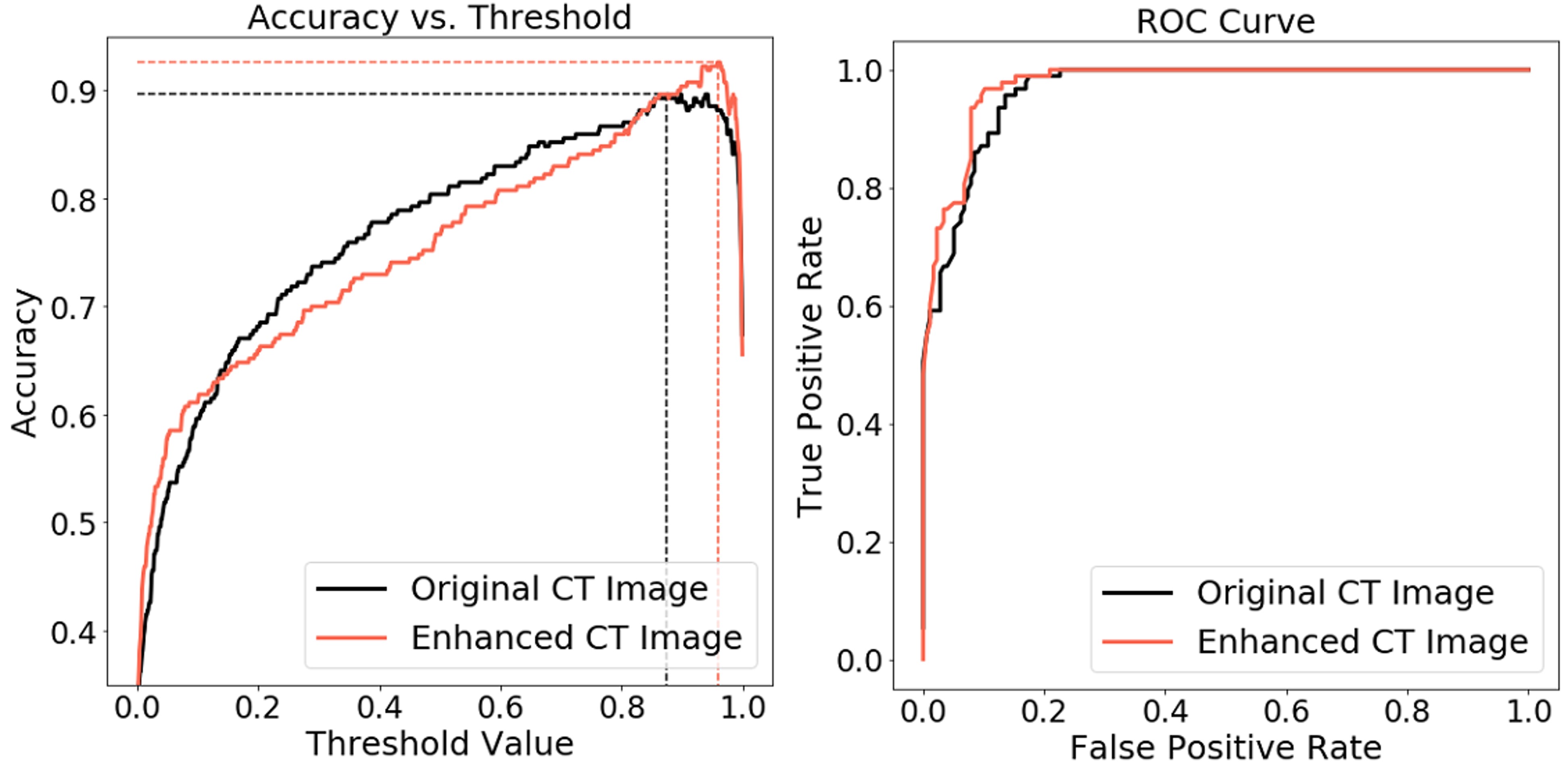}
        \caption{\textbf{(a)} Diagnostic accuracy in different thresholds. \textbf{(b)} ROC curve changing of classification sub-model between original CT image based and enhanced CT image based.}
    \label{fig:2}
\end{figure}

\section*{Tables}

\begin{table}[H]
\centering
\caption{Optimal values of hyper-parameters for training 2D-DDnet}
\label{tab:train_param2d}
\begin{tabular}{cc}
\toprule[2pt]
Parameter & Value \\ \hline
Epochs                                   & 50             \\ 
Learning rate                            & 0.0001         \\ 
Learning rate decay                & 0.95            \\ 
Batch size                               & 2              \\ \bottomrule[2pt]
\end{tabular}
\end{table}

\begin{table}[H]
\caption{Optimal values of hyper-parameters for training 3D-DDnet}
\label{tab:train_param3d}
\centering
\scalebox{1}{
\begin{tabular}{cc}
\toprule[2pt]
Parameter  & Value \\ \hline
Epoch               & 60                                  \\ 
Learning rate       & 0.0001                               \\ 
Learning rate decay & 0.95                                \\ 
Batch size          & 2                                   \\ \bottomrule[2pt]
\end{tabular}}
%\vspace{-15pt}
\end{table}

\begin{table}[H]
\centering
\caption{Hyper-parameters for training classification AI}
\label{tab:cla_train_para}
\begin{tabular}{cc}
\toprule[2pt]
Parameter                              & Value                             \\ \hline 
Optimizer                                   & ADAM                               \\ 
Learning Rate                               & $10^{-6}$         \\ 
Gaussian Noise                              & probability = 0.75, variance = 0.1 \\
Image Contrast                              & probability = 0.5                  \\
Image Intensity Oscillation Scale Magnitude   & $10^{-1}$         \\ \bottomrule[2pt]
\end{tabular}
\end{table}

\begin{table}[H]
\centering
\caption{Quantitative analysis of image enhancement by the 2D-DDnet
%Y and X refers to high dose and low dose CT images. f(x) is the image enhanced by DDnet.
}
\label{tab:ddnet_quantitative}
\begin{tabular}{ccc}
\toprule[2pt]
       &MSE & MS-SSIM \\ \hline
Low quality CT images    & $0.00715 \pm 0.00724$                          & $96.2 \%  \pm 2.1\%$                             \\ 
Enhanced CT images & $0.00098 \pm 0.00082$                          & $98.6 \%  \pm 1.1\%$                              \\ \bottomrule[2pt]
\end{tabular}
\end{table}

\begin{table}[!ht]
\centering
\caption{Quantitative analysis of volumetric enhancement by the 3D-DDnet.  
%Y and X refer to high-dose and low dose CT images.  f(x) is the image enhanced by DDnet
}
\label{tab:3d_enhancement_quant}
\begin{tabular}{cccc}
\toprule[2pt]
      Scan Size& Comparison against high quality CT images & MSE & 3D MS-SSIM \\ \hline
                                           & Low quality CT images                 & $0.01142 \pm 0.0059$                          & $89.50\% \pm 3.2\%$                                \\ 
{\multirow{-2}{*}{$512\times512\times64$}}                & Enhanced CT images              & $0.00309 \pm 0.0069$                          & $92.10\% \pm 4.1\%$                                \\
\bottomrule[2pt]
\end{tabular}
\end{table}

\begin{table*}[!ht]
\centering
\caption{Quantitative comparison of image and volumetric enhancement using DDnet. The data reported correspond to 2D and 3D version of DDnet trained and tested on same data.}
\label{tab:comp_2d_3d_ddnet}
\scalebox{0.9}{
\begin{tabular}{ccccc}
\toprule[2pt]
Network & Scan Size                                       & Comparison against high quality CT images & MSE & 3D MS-SSIM \\ \hline
\multirow{2}{*}{2D-DDnet}              & \multirow{2}{*}{$512\times512\times64$}                                                     & Low quality CT images                  & $0.01142 \pm 0.0059$                            & $89.50\% \pm 3.2\%$                                \\ 
                                       &                                                                                 & Enhanced CT images              & $0.00743 \pm 0.00041$                            & $91.40\% \pm 1.8\%$                                \\ %\cline{2-5} 
                                      % & \multirow{2}{*}{$512\times512\times16$}                                                     & Y-X                 & 0.0109                            & 88.3\%                                \\ 
                                       %&                                                                                 & Y-f(x)              & 0.0067                            & 90.8\%                                \\ 
                                       \hline
\multirow{2}{*}{3D-DDnet}             % & \multirow{2}{*}{$256\times256\times64$}                                                     & Y-X                 & 0.0107                            & 94.0\%                                \\  
                                      % &                                                                                 & Y-f(x)              & 0.0013                            & 98.6\%                                \\ \cline{2-5} 
                                       & \multirow{2}{*}{\begin{tabular}[c]{@{}c@{}}$512\times512\times64$\\ (4 Tiles)\end{tabular}} & Low quality CT images                 & $0.01142\pm 0.0059$                            & $89.50\% \pm 3.2\%$                                \\ 
                                       &                                                                                 & Enhanced CT images              & $0.00309 \pm 0.0069$                            & $93.1\%\pm 4.1\%$                                \\ %\cline{2-5} 
                                     %  & \multirow{2}{*}{$512\times512\times16$}                                                     & Y-X                 & 0.0109                            & 88.3\%                                \\ 
                                       %&                                                                                 & Y-f(x)              & 0.0034                            & 92.1\%                                \\ 
                                       \bottomrule[2pt]
\end{tabular}
}
\end{table*}

\begin{table}[H]
\centering
\caption{Statistical analysis of the 2D CT image diagnosis simulation results}
\label{tab:statResult}
    \begin{tabular}{cccc}
    \toprule[2pt]
                       & Original CT Image & Enhanced CT Image & Percent Increase \\ \hline 
    % \textbf{Positive APPS} & 0.4364                     & 0.5500                    & 26\%               \\
    Precision & 0.7876                     & 0.8613                    & 9.37\%               \\
    Recall    & 0.9570                     & 0.9355                    & -2.25\%              \\
    F1-score  & 0.8641                     & 0.8969                    & 3.80\%               \\
    Accuracy  & 0.8963                     & 0.9259                    & 3.31\%               \\
    AUC       & 0.9653                     & 0.9748                    & 0.99\%               \\ 
    \bottomrule[2pt]
    \end{tabular}
\end{table}

\end{document}